\documentstyle[amssymb,aps]{revtex}

\tightenlines

\begin{document}
\title{Anisotropy of x-ray scattering in aligned nanotube structures}
\author{A.V. Okotrub }
\address{SB RAS - Nikolaev Institute of Inorganic Chemistry, 630090\\
Novosibirsk, Russia}
\author{S.B. Dabagov\thanks{%
Author to whom all correspondence should be addressed; electronic mail:
dabagov@lnf.infn.it}}
\address{INFN - Laboratori Nazionali di Frascati, I-00044 Frascati (RM),\\
Italy \\
and RAS - P.N. Lebedev Physical Institute, 119991 Moscow, Russia}
\author{A.G. Kudashev, A.V. Gusel'nikov, L.G. Bulusheva}
\address{SB RAS - Nikolaev Institute of Inorganic Chemistry, 630090\\
Novosibirsk, Russia}
\author{A. Grilli}
\address{INFN - Laboratori Nazionali di Frascati, I-00044 Frascati (RM),\\
Italy}
\maketitle

\begin{abstract}
Effects of orientational x-ray scattering have been experimentally examined
in a film of vertically aligned multiwall carbon nanotubes (CNs). Additional
contribution to the x-ray fluorescence intensity was revealed at angles
close to the film normal. Theoretical considerations suggest the intensity
enhancement to be caused by propagation of C K$_{\alpha }$-radiation mainly
along the channels of CNs.
\end{abstract}


One of the attractive features of CN morphology is a long inner channel \cite%
{sa...book98}, which, in particular, could be used to handle x rays,
neutrons and charged particle beams \cite%
{klle-pla96,ded-nimb98,zhgl-pla98,kru...apl03}. As well known, x rays and
thermal neutrons can be effectively transported by a hollow part of
microcapillaries, and that opens an opportunity for the creation of new
optical elements \cite{kuko-phrep90,ku-spie00}. In addition to the high
practical importance, capillaries have been shown to represent a fine tool
for studying fundamental features of radiation scattering by a surface \cite%
{cada...apl01}. In microcapillaries the size of hollows is much larger than
the effective wavelength of the radiation and the radiation dispersion can
be described within the framework of surface channeling. In case of CNs, the
effective wavelength of radiation becomes comparable with a diameter of
channels and instead of surface scattering effects we deal with bulk
features of radiation interaction with a nanotube structure \cite{da-phusp03}%
. The aligned CNs are attractive for understanding the nature of radiation
propagation and scattering in the channels of extremely small diameters. The
present work reports the first experimental study of the influence of
nanotube alignment on the angular dependence of x-ray fluorescence.

Obviously, measurements of x-ray scattering in an individual CN are just
provisional and such type of experiments could be first performed for a bulk
material consisting of the ordered nanotubes. Nowadays, a number of
techniques to produce the films of oriented nanotubes have been developed.
Mostly aligned multiwall CNs have been synthesized using a zeolite-supported
catalyst \cite{li...sci96}, the lithography technique \cite%
{solee-apa02,te...chpl98}, and mixtures of ferrocene and fullerene or
hydrocarbon compounds \cite{may...chpl01,gro...apl99,sat...chpl99}. Both the
nanotube structure and the character of its ordering are varied as a
function of the synthesis conditions.

As a starting point, a film of vertically aligned nanotubes, grown on a Si
substrate, was prepared by thermal chemical vapor deposition of a
ferrocene/fullerene mixture. The chemical vapor deposition apparatus
consists of a gas flow reactor and a tubular furnace; the reactor is fitted
with a removable quartz tube. A ceramic boat with a mixture of fullerene C$%
_{60}$ and ferrocene Fe(C$_{5}$H$_{5}$)$_{2}$, taken in the ratio of 1:1,
was placed inside the tube beyond a Si 4$\times $10 mm$^{2}$ plate. The
pyrolysis was performed in an argon flow of atmospheric pressure at 950$^{o}$%
C \ \cite{ok...conf03}. The material produced was characterized by a
scanning electron microscope (SEM). Figure 1 shows a SEM image of a CN
layer. The mean length of nanotubes coincides with the film thickness of
about 10 $\mu $m. The nanotubes are rather closely packed and their density
was estimated about 1000 pieces/$\mu $m$^{2}$. A transmission electron
microscope showed that the multiwall CNs have from 5 up to 15 layers, the
external diameter of $\sim $ 10 $\div $ 20 nm and the inner channel diameter
of $\sim $ 4.0 $\div $ 6.0 nm. From the SEM analysis it follows that the
nanotubes are mostly ordered for an internal part of the film. In the
beginning of the growth process the silicon surface with deposited iron
particles (as the catalyst), provides the alignment of CNs. An upper part of
the film (0.1 $\div $ 0.2 $\mu $m thickness) is the most disoriented due,
perhaps, to the final growing features, when the concentration of fullerenes
is lower.

The angular dependence of x-ray fluorescence from nanotube films was
measured by means of a laboratory spectrometer. A scheme of the sample
location within a spectrometer chamber is presented in Fig. 2. The operating
vacuum was 2$\times $10$^{-4}$ Pa. In order to change the angle $\alpha $
between the sample surface and the optical axis of spectrometer from 60$^{o}$
to 100$^{o}$, the sample of 8$\times $2 mm$^{2}$ was fixed to the rotatable
axis. Positions of the Cu anode of x-ray tube and the entrance slit were
fixed. The angle between the radiation source, i.e. the anode surface of 6
mm width, and the rotating axis was 20$^{o}$. A gas proportional counter
with methane pressure of 0.2$\times $10$^{5}$ Pa was placed beyond the slit.
A counter window was made from polypropylene film of 2 $\mu $m thickness.
Maximal counter charge was $\sim $ 3000 impulses/sec. at anode voltage of
2.3 kV and current of 100 mA; the total intensity at maximum accounted $\sim 
$ 60000 impulses.

To reveal channeling effect in CNs, x-ray fluorescence from the aligned and
randomly distributed nanotubes should be compared. Random distribution of
nanotubes was obtained by applying pressure to the nanostructure film. The
complete disorientation of nanotubes with respect to the substrate surface
was confirmed by the SEM analysis. The angular dependences of x-ray
fluorescence measured for the samples of ordered as well as chaotic nanotube
distributions are shown in Fig. 3 (curves 1 and 2, respectively). The count
values were normalized by the maximum. The curves show similar intensity
dependence: increase at angles from 60$^{o}$ up to 85$^{o}$ and then
decrease. The sample of aligned CNs provides an additional contribution to
the x-ray fluorescence spectrum near 90$^{o}$. For the sample of disordered
nanotubes the angular dependence will be defined by geometrical parameters
of the experimental layout and by absorption of the exciting and
fluorescence beams. The measured intensity was fitted by the following
expression: $I=I_{0}\sin \alpha \sin (110^{o}-\alpha )\exp \left( -\frac{%
\eta \cos (\alpha -20^{o})}{\sin \alpha }\right) $, \noindent where I$_{0}$
is the initial radiation intensity, $\eta $ is the cumulative factor
determined by the absorption. The formula takes also into account the change
of the acceptance vs the angle $\alpha $. This dependence is shown by the
curve 3 in Fig. 3, and it has a maximum at 82$^{o}$. The simulated angular
dependence for x-ray fluorescence has good agreement with the measured one
for the disordered nanotube film and shows absence of peculiarities in the
range of angles close to the normal.

Figure 3 presents also the angular distribution of x-ray fluorescence from
the Si substrate (curve 4). Due to the presence of a SiO$_{x}$ layer on the
substrate surface, a fluorescence spectrum is formed by the O K$_{\alpha }$,
Si K$_{\alpha }$, Si K$_{\beta }$ and Si L$_{2,3}$ lines. As seen, the
angular dependences for the substrate and nanotube film are rather similar,
while in the case of clean substrate the integral intensity is 5 times less.
The reason of low intensity can be due to comparative low probability for Si
ionization ($\omega _{\text{ion}}$ = 1.8 keV) by the bremsstrahlung of x-ray
tube. This, in turn, leads to low intensity of the both Si K$_{\alpha }$ and
K$_{\beta }$ lines. Contribution of x-ray emission by oxygen could be
comparable with the C K$_{\alpha }$ spectra of CNs, however, its quantity is
small. The Si L$_{2,3}$ lines signal detected by the gas proportional
counter is very weak because of relatively large thickness of the counter
window. The decrease in fluorescence emission from the substrate (curve 4)
with respect to the x-ray emission in presence of the CN film (curves 1 and
2) at scattering angles less than 70$^{o}$ is due to the change in the
absorption coefficients for Si and C. Thus, even the substrate fluorescence
contributes in the forming of the curves 1 and 2, its relative value is
rather small, and we can neglect it.

A five-time enlarged difference between the curves 1 and 2 is given by the
curve 5. The integral presented by the curve 5 was estimated to be $\sim $
0.4 \% of the total fluorescence intensity. Despite the large statistical
spread, this dependence obviously exhibits a maximum of $\sim $ 6$^{o}$ in
width at 90$^{o}$ that based on "a plato" of $\sim $ 20$^{o}$ in width. We
suppose that this feature is caused by the influence of CNs alignment on
x-ray fluorescence, and the maximum width is mainly defined by the deviation
of CNs from a vertical alignment. X-ray fluorescence from strongly bent CNs
that takes place near the film surface, gives a broad diffuse halo. The
distribution 5 is determined by the combination of various mechanisms of
radiation scattering in the nanotube film. We shall try to evaluate
contribution by the main scattering factors in the total angular
distribution of radiation behind the studied structures.

Mean angular deviation $\overline{\Delta \theta }$ of x-ray fluorescence
after traversing a CN film can be estimated, paying attention to the fact
that, originally, x-ray fluorescence is isotropic. Obviously, the width is
defined by coherent and incoherent radiation scattering in the inner cavity
and the walls of nanotubes (the latter can be considered as radiation
diffraction on a multilayer system) $\overline{\Delta \theta _{ch}}$, as
well as by diffuse scattering in the space between separate nanotubes and
incoherent scattering by the structure imperfections $\overline{\Delta
\theta _{d}}$, i.e. $\overline{\Delta \theta ^{2}}=\overline{\Delta \theta
_{d}^{2}}+\overline{\Delta \theta _{ch}^{2}}$. \noindent In turn, the
diffuse term depends on the texture of nanotube layer and on the mean
distance between the separate nanotubes or the bundles of nanotubes tightly
attached to each other, i.e. $\overline{\Delta \theta _{d}^{2}}=\overline{%
\Delta \theta _{t}^{2}}+\overline{\Delta \theta _{p}^{2}}$. Incoherent
scattering due to the film texture $\overline{\Delta \theta _{t}}$ is
determined by radiation absorption in the tube walls and by imperfections,
while the second term $\overline{\Delta \theta _{p}}$, which corresponds to
the free fly-by (straight ray) propagation, can be easily evaluated from the
geometrical structure parameters.

Detailed examination of microscopic pictures of the samples allows the
texture parameters to be defined. Namely, our analysis shows that for a film
thickness $l\approx $ 10 $\mu $m, the mean distance between the separate
nanotubes or the bundles of nanotubes is about $a\approx $ 100 nm. More than
that, mean angular deviation in the angular orientation for the nanotubes
layer is in the range of 30$^{o}$ $\div $ 50$^{o}$ that defines $\overline{%
\Delta \theta _{t}}$. For the studied samples the term $\overline{\Delta
\theta _{p}}\simeq $ $\frac{a}{l}$, corresponding to the straight ray
propagation between the nanotube bundles, does not exceed a few mrads ($%
\lesssim $ 1$^{o}$). This effect, so called \textquotedblleft collimation
effect\textquotedblright , strongly depends on the depth of radiation source
position; due to the low density of the film, any nanotube atom can be a
source of x-ray fluorescence \cite{zhe...83}. As well known, the C K$%
_{\alpha }$-spectrum has an energy extended structure in the range of
275-284 eV close to the K-edge ($\approx $ 285 eV), and the coefficient of
absorption is rather low for the photons near the absorption edge \cite%
{he...adndt82}. Thus the effect of collimation will be suppressed. Moreover,
we can conclude that diffusive part of radiation scattering in the nanotube
film forms a wide background in the angular distribution of x-ray
fluorescence.

Now let us evaluate $\overline{\Delta \theta _{ch}}$ that we defined as a
term describing radiation scattering inside aligned channels (nanotubes). In
this case the problem becomes similar to the electromagnetic wave
propagation in optical fibers \cite{glo-phrep79,zhegle-nima93}. Following a
general definition of scattering at channeling (see \cite{da-reds93-1}),
mean-square angle of radiation scattering in a nanotube can be estimated by
: $\overline{\Delta \theta _{ch}^{2}}\propto \int\limits_{\omega _{\min
}}^{\omega _{\max }}d\omega \ \frac{N_{ph}(\omega )}{\omega ^{2}}%
\int\limits_{0}^{r_{_{0}}}dx\ N_{e}(x)$, \noindent where $r_{0}$ is the
internal radius of empty cylindrical nanotube core, $\omega $ and $%
N_{ph}(\omega )$ correspond to the emitted photon energy and the number of
photons with energy $\omega $, $N_{e}(x)$ is the electron density and $x$ is
the radial coordinate in the transverse nanotube cross-section \cite%
{ded-nimb98,zhgl-pla98}. Reducing this expression results in $\overline{%
\Delta \theta _{ch}}\approx 3^{o}\div 6^{o}$ at the given values of $\omega
_{0\max }$$\sim $ 100-170 eV and $r_{0}$$\sim $ 2-3 nm. However, this
estimation is valid just for the proper channeling regime (an ideal case),
when the effects of dechanneling (when x-ray beams leave nanotube channels)
are suppressed. In the considered case we deal with multiwall bent
nanotubes. And first, it is necessary to take into account dechanneling of
photons due to the strong nanotube bending, at which the requirements for
x-ray channeling are violated . Indeed, essential bending through the short
distances causes increasing the angles of radiation propagation with respect
to the nanotube longitudinal axis, that results in fast dechanneling of x
rays. Thus, x-ray channeling takes place just in the rather straight parts
of nanotubes. However, these parts are governed by the film texture. Due to
that, the angular distribution formed by x rays at channeling could be wide.
Herewith, studying the SEM images by the Fourier analysis has shown that on
the base of resolvable nanotube disorientation, one can observe a sharp
narrow (5$^{o}\div $10$^{o}$) peak for nanotubes normally directed \cite%
{kin...conf03}.

What deserves special mention is the tunneling through the potential barrier
that in our case is responsible for the radiation leakage through the
nanotube wall. In our samples, the thickness of walls $\Delta d$ varied in
the limits of $\ $$\sim $ 3 $\div $ 7 nm, which is of the order of the
transverse wavelength of radiation $\lambda _{\bot }\,\gtrsim $ 4.42 nm.
Therefore, the probability of tunneling , $\propto \exp \left( -\frac{\Delta
d}{\lambda _{\bot }}\right) $, is quite high. That's probably why relative
integrated power of the narrow peak in the angular distribution of emerged
radiation is very weak. Moreover, the tunneling process is accompanied by a
small-angle diffraction. Presence of a set of equidistant layers in the wall
leads to the diffraction dispersion of x-ray photons in such a manner that
the first diffraction maximum is practically observed at angles close to the
scattering angles under channeling ($\sim $ $5^{o}$).\newline

In our work we studied orientational effects of x-radiation scattering in a
layer of nanotubes. During measurements of x-ray fluorescence from a film of
aligned multiwall CNs, the intensity amplification was revealed at angles
close to the normal. The width of a reflection curve is mainly defined by
the angular deviation in orientation of nanotubes.

Recent development of the technology of nanostructure manufacturing
testifies more and more to the opportunity for cultivation of layers of
aligned nanotubes. Thus, the question on the use of hollow nanostructure
channels becomes rather actual. It is obvious, that the quality of film
arrangement is still far from perfect in order to consider these structures
as objects with the stable allocated directions. However, since the
discussed effects can be observed on micron distances due to extremely small
channel sizes, already today, it is possible to start studying structural
effects of radiation scattering in CNs.

\newpage%

\section*{List of figures}

\noindent Figure 1: SEM image of a CN layer grown on a Si substrate.

\noindent Figure 2: Scheme of sample (S) location in the x-ray tube relative
to exciting (II) and fluorescence beams (I).

\noindent Figure 3: Angular dependence of x-ray fluorescence measured for
the film of aligned (curve 1) and randomly distributed (curve 2) CNs. Curve
3 presents the fitted angular dependence of x-ray fluorescence. Curve 4
corresponds to the angular distribution of x-ray fluorescence of the clean
Si substrate. The five-time multiplied difference between curves 1 and 2 is
shown by curve 5.

\bigskip

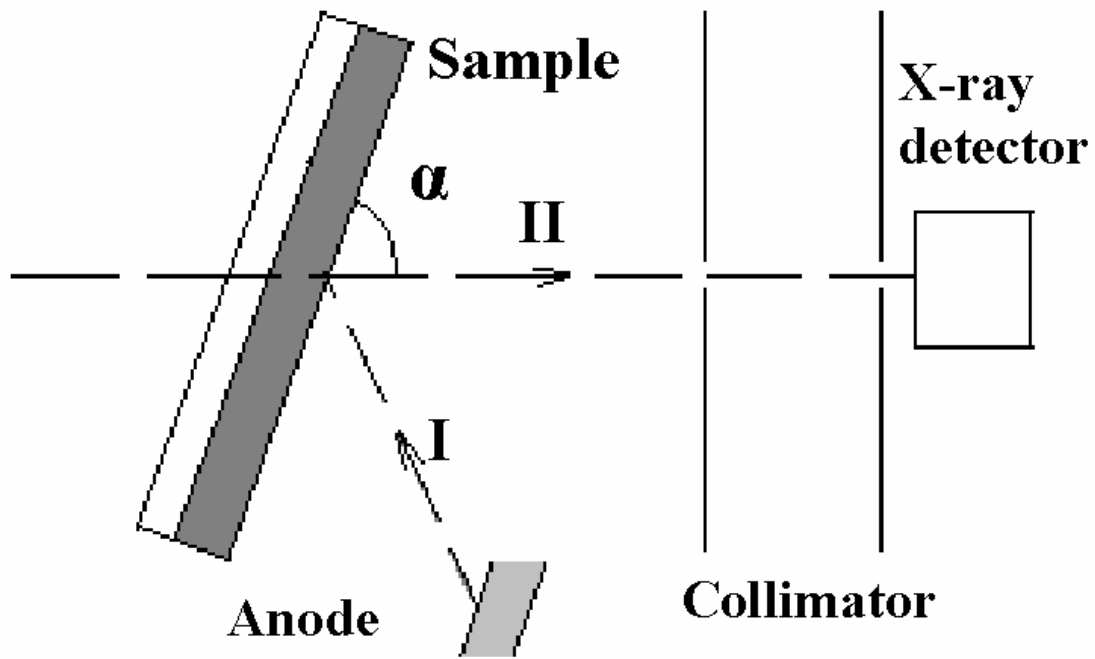
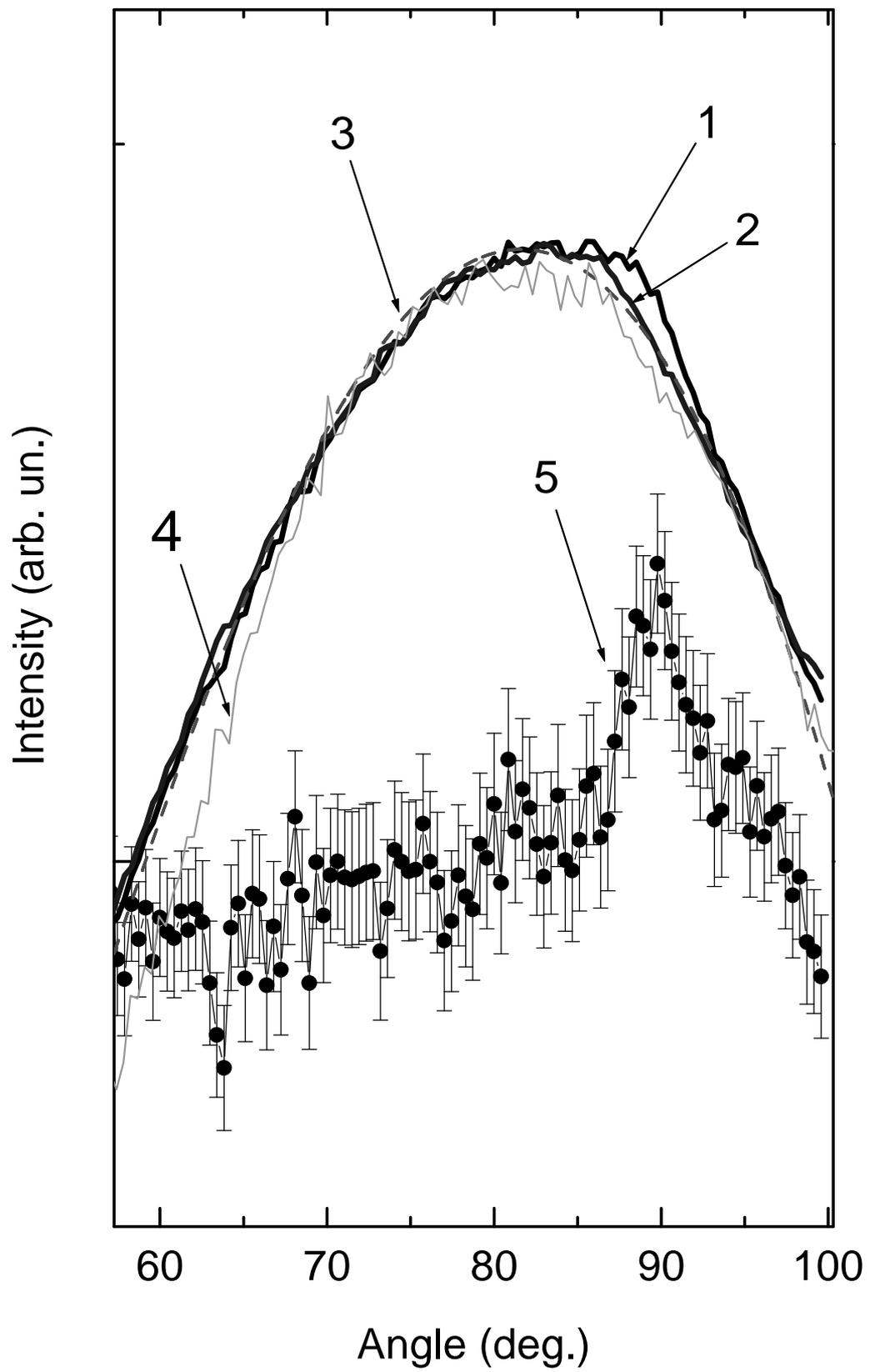
\end{document}